\documentclass[conference,compsoc]{IEEEtran}

\usepackage{cite}
\usepackage{graphicx}
\usepackage{titlesec}

\usepackage{amsfonts} 
\usepackage{amsmath}
\usepackage{amsthm}
\usepackage{dsfont}

\usepackage{tikz} % the tikz package  
\usetikzlibrary{shapes.geometric,arrows} % to implement the corresponding shapes and arrows, these are used.  
\usetikzlibrary{positioning}

\usepackage{caption}
\usepackage{subcaption}

\usepackage{hyperref}
\usepackage[T1]{fontenc}
\usepackage{xcolor}
\usepackage{smartdiagram}

\theoremstyle{definition}

\theoremstyle{ass}

\begin{document}

\title{Robustness of the  Tangle 2.0 Consensus}

\author{
\IEEEauthorblockN{working on it at some point or another:  Bing, Daria, Piotr, Andreas, Sebastian, Gal \\ 
interested: Billy, Jonas}
\IEEEauthorblockA{IOTA Foundation\\
Berlin, Germany\\
Email: research@iota.org}
}

\author{
\IEEEauthorblockN{Bing-Yang Lin\IEEEauthorrefmark{1},
Daria Dziubałtowska\IEEEauthorrefmark{1},
Piotr Macek\IEEEauthorrefmark{1},
Andreas Penzkofer\IEEEauthorrefmark{1},
Sebastian Müller\IEEEauthorrefmark{1}\IEEEauthorrefmark{2}}
\IEEEauthorblockA{\IEEEauthorrefmark{1}IOTA Foundation, 
Berlin, Germany}
\IEEEauthorblockA{\IEEEauthorrefmark{2}Aix Marseille Universit\'e, CNRS, Centrale Marseille, I2M - UMR 7373,  13453 Marseille, France}}
\maketitle

\maketitle
\thispagestyle{plain}
\pagestyle{plain}

\begin{abstract}
In this paper, we investigate the performance of the Tangle 2.0 consensus protocol in a Byzantine environment. We use an agent-based simulation model that incorporates the main features of the Tangle 2.0 consensus protocol. Our experimental results demonstrate that the Tangle 2.0 protocol is robust to the bait-and-switch attack up to the theoretical upper bound of the adversary's $33\%$ voting weight. We further show that the common coin mechanism in Tangle 2.0 is necessary for robustness against powerful adversaries. 
Moreover, the experimental results confirm that the protocol can achieve around $1$s confirmation time in typical scenarios and that the confirmation times of non-conflicting transactions are not affected by the presence of conflicts.


\end{abstract}

\IEEEpeerreviewmaketitle

\section{Introduction}

Since 2009 when Bitcoin \cite{nakamoto2008bitcoin} was first introduced, distributed ledger technologies (DLTs) have gained growing interest from academics, industries, and even governments \cite{systematicReview}. In DLTs, ledger records have no need for central authority controls or maintenance, and the validity of each record relies on a decentralized consensus mechanism, or globally accepted truth between participants, also called nodes, in the network. In Bitcoin, a blockchain is adopted as the structure to store a ledger, Proof of Work (PoW) as the Sybil protection, and the longest chain rule as the consensus mechanism. Each block, which contains transactions, is linked in a linear growing chain. All nodes in the Bitcoin network are able to extend the chain  by solving a numerical challenge that consumes a considerable amount of computation power. The node that solves the challenge first is elected as a ``leader'', can add the next block to the chain, and gain a reward. Although  Bitcoin was revolutionary in enabling DLTs in the first place, the PoW mechanism is typically slow \cite{scalingByzantineConsensus} and might demand an unsustainable energy consumption \cite{consensusProtocols}. Also, the single chain structure of the Bitcoin ledger limits its throughput \cite{ACCEL}, and has a scalability problem \cite{blockchainScalability}.

As a consequence, instead of using a linear block chain to store the ledger, many non-linear ledger structures (e.g., SPECTRE \cite{spectre}, Byteball \cite{byteball}, Algorand \cite{gilad2017algorand}, PHANTOM \cite{phantom}, Avalanche \cite{Ava19}, Prism \cite{bagaria2019prism}, AlephZero \cite{gkagol2019aleph}, Narwhal \cite{Narwhal22}, and IOTA \cite{2020coordicide}) were proposed to improve the performance. 
The consensus mechanism in a DAG-based system can be conceptually different from the one in a linear blockchain system, and the transaction throughput is potentially no longer limited. In \cite{dagReview}, the scalability and efficiency of different DAG-based blockchain architectures were analyzed, based on their functional data structures. We also refer to \cite{sok} for an overview of the security and performance of several DAG-based blockchain systems. 

\subsection{Results and contribution}
A recent proposal for a DAG-based consensus protocol is the Tangle 2.0 protocol, \cite{OTV}. Our paper builds on the theoretical foundations and mathematical models established in  \cite{OTV} and gives a first performance analysis of the protocol in a Byzantine environment. To this end, we provide an agent-based simulator, \cite{otv-simulator}, that simulates the peer-to-peer layer, the leaderless block creations, and the consensus finding simultaneously. In contrast to previous DAG simulators, such as DAGSim \cite{DagSim}, our simulator \cite{otv-simulator} allows to consider a Byzantine environment and incorporates the main features of the Tangle 2.0 consensus. 

This work is the first paper that provides quantitative validations of the Tangle 2.0 consensus protocol. In addition, we show that confirmation time is in the order of a second. This good performance holds even in highly adverse environments, where attackers are effective up to their theoretical limits. 

The Tangle 2.0 consensus protocol consists of two components: the asynchronous component, On Tangle Voting (OTV), and the synchronous part, Synchronised Random Reality Selection (SRRS). To explore the security of these components, an agent-based attack strategy, called Bait-and-Switch, is introduced in the simulations. In this kind of attack, adversaries can issue double-spends at arbitrary high frequency and aim to keep the honest nodes in an undecided state. 
The simulation results show that in the worst weight distribution case, where all the honest nodes have equal weight, the OTV protocol can still resist the Bait-and-Switch attack when the adversary node owns up to $20\%$ of the total weight. In addition, we also show that the SRRS protocol can further resist the Bait-and-Switch attack, even when the adversary occupies up to $33\%$ of the total weight, which is the theoretical upper bond.

\subsection{Structure of the paper}
In Section \ref{sec:tangle}, we introduce the fundamentals of the Tangle 2.0 protocol. In Section \ref{sec:attack}, we define the adversary model and describe the Bait-and-Switch strategy. In Section \ref{sec:simulation}, we describe the components considered in our simulations and explain the setup in Section \ref{sec:setup}. The experimental results are given in \ref{sec:results}, and a conclusion can be found in Section \ref{sec:conclusion}.

\section{Fundamentals of the Protocol} 

We introduce several fundamental concepts relevant to the protocol in our context. First, we discuss the UTXO accounting model, before addressing the block DAG that, in combination with a suitable Sybil protection facilitates the voting schemes.

\subsection{The UTXO Ledger and Conflicts}
Tracking of funds and change of ownership between addresses is facilitated by employing the  Unspent Transaction Output (UTXO) model, e.g., \cite{nakamoto2008bitcoin, cardanoEUTXO, UTXO}. In this model,  transactions specify the outputs of previous transactions as inputs and spend them by creating new outputs. In contrast to the account-based model, transactions can be verified without knowing the global state of the system but depend only on the status of the inputs. Moreover, it identifies conflicts as every output can only be spent once. 

As consistency is the main requirement of a ledger, nodes eventually have to resolve this kind of double-spends. The Tangle 2.0 consensus protocol, \cite{OTV}, decides between the conflicting spending relying on an identity-based Sybil protection, Section \ref{sec:reputation}, and a voting protocol introduced in more detail in Sections \ref{sec:AWVoting} and \ref{sec: SRRS}.

\subsection{The Tangle}\label{sec:tangle}

One of the basic ideas of \cite{theTangle}  was to abolish the role of the miners or validators, as a leader-based approach not only leads to a performance limitation but also makes fees necessary. Instead, a cooperative approach is followed, in which each new block validates at least two other previous blocks by referencing them.
Each participant can thus add blocks simultaneously and asynchronously. The blocks together with the references then form a DAG, the so-called \textit{Tangle}. 
The Tangle is both a record of the communication between nodes as well as  facilitates a data structure for the voting schemes introduced in the next sections. 

\begin{figure}
    \centering
    \includegraphics[width=0.45\textwidth]{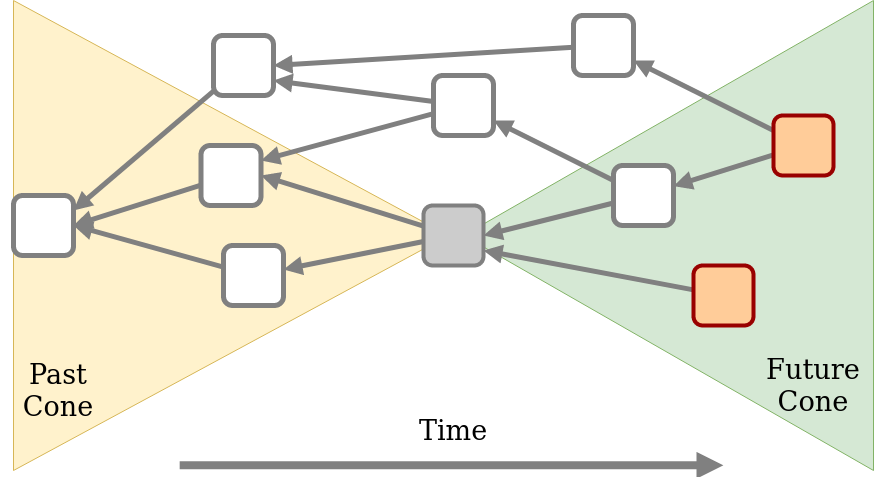}
    \caption{The Tangle, the future, and the past cone of a block are highlighted with green and yellow triangles respectively. The orange blocks are tips.}
    \label{fig:tangle}
\end{figure}

We call the referenced and approving blocks \emph{parents} and \emph{children}, respectively. The reference relationship is transitive in the sense that new nodes not only refer (or vote) for their direct parents but also indirectly for the entire ancestral line. We call all blocks, referenced directly or indirectly, the  \emph{past cone} of a given block. Similarly, all blocks that refer to the block directly or indirectly are called its \emph{future cone}, see Figure \ref{fig:tangle}. 
The first block of the Tangle  has no parents and is  called \emph{genesis}.   Blocks that have not been yet approved are called \emph{tips}.
In \cite{OTV} there exist different reference types that infer different past cones of a block. However, in this paper, we consider only the most fundamental reference type which represents approval for the block itself and its entire past cone of the Tangle.

Every block consists of several elements, of which the most relevant are: parent references, the signature of the issuer (the node that introduced the block into the Tangle), and a payload. In this paper, the payload is a transaction.

\subsection{Sybil Protection}
\label{sec:reputation}
Each node has a score, called \textit{weight}, which serves as a Sybil protection mechanism. 
Weights can be calculated in different ways, depending on the purpose. For example, it could be utilized in the context of (delegated) Proof-of-Stake systems. In the context of this paper, we employ it as voting power and access to the network's resources (share of the network throughput).

The total weight value within the network is the sum of the weights of all nodes. We note, that in a distributed network setting nodes can go offline and online at any time. Also, nodes may not be  forced to issue blocks and use all of their bandwidth. However, in this work, we assume that the weights of the nodes are constant and that all nodes use their total share of the network throughput.

\subsection{Approval Weight, Voting, and Confirmation}\label{sec:AWVoting}

In this section, we define the concept of Approval Weight (AW). It is the core element of the consensus protocol. AW allows for measuring the endorsement of blocks and their corresponding transactions.  The confirmation of a transaction is calculated by summing up the cumulative weight of all blocks' issuers in the future cone of the block that contains the transaction. It is in principle similar to the concept introduced in \cite{theTangle}. There, a weight of a block is corresponding to the amount of work put into creating a block and all blocks in its future cone. Instead, in this implementation block weight is equal to the issuer's weight and is counted at most once if a block is issued in the future cone. A block becomes confirmed whenever its accumulated weight reaches a certain threshold $\theta$.

The On Tangle Voting (OTV), \cite{OTV}, allows voting on two conflicting transactions or double spends, $A$ and $B$.
A given node might receive transaction $A$ and express its support for it (by issuing a block in the future cone of the block containing $A$) before receiving $B$. Once the node also receives $B$ and its future cone, it compares each transaction's AW and votes for the transaction with the higher weights. In case the node changes its opinion in favor of $B$, the node's weight is revoked from the AW of transaction $A$, since for any pair of conflicts a node's weight is counted at most once. 

Once a transaction is confirmed, i.e., its AW reaching $\theta$, it will stay confirmed even if its AW falls below $\theta$.
Thus, the consensus protocol must be designed, such that the event of any two nodes confirming different transactions out of a set of mutually conflicting transactions occurs with negligible probability, or otherwise safety as discussed in the next section, is violated.

\subsection{Liveness and Safety}

Every consensus algorithm should ensure the following properties:
\begin{itemize}
 \item \textbf{liveness}: all non-faulty nodes eventually take a decision,
    \item \textbf{safety}: all non-faulty nodes agree on the same values.
\end{itemize}
The  satisfaction of the two properties depends on the underlying communication model. The most well-known communication models are synchronous, asynchronous, and partial synchronous, describing the network's synchronicity level. 

The asynchronous model best reflects scenarios where an adversary can delay the transmission of blocks arbitrarily long. From the famous FLP impossibility result, \cite{fischer1985impossibility}, we know that reaching a (deterministic) agreement in an asynchronous setting is impossible. Therefore, there is a need for a so-called symmetry-breaking mechanism that uses a (shared) source of randomness to ensure liveness and safety.

It is crucial that the safety property always holds (at least with high probability). It is also essential that the progress of the protocol should happen eventually.
The main goal of the performed simulations is to determine if it is possible to halt the termination of the voting mechanism and if the proposed symmetry or metastability breaking mechanism, described in the next section, ensures both the safety and liveness of the protocol can be guaranteed.

\subsection{Synchronised Random Reality Selection}\label{sec: SRRS}

The concept of Synchronised Random Reality Selection (SRRS) was proposed in \cite{OTV} and inspired by \cite{POPOV2021}. Because the OTV is an asynchronous protocol, a synchronization process on the communication level is lacking between nodes to overcome the deterministic FLP impossibility result. 
This synchronization of the nodes is done in the SRRS using  a shared random number that is generated periodically. We denote  $D$ as the period (epoch) time.
By the end of each epoch, every node in the network receives the newly generated random number. This ``common coin'' is used to \emph{synchronize} the opinions of each node with the opinions of the other honest nodes. 
More details on the SRRS are described in Section \ref{sec:srrs-implementation}.

\section{The Adversary Model}
\label{sec:attack}

 We distinguish between two types of nodes: \textit{honest} and \textit{malicious}. Honest nodes follow the protocol, while malicious nodes are trying to actively disturb the protocol by not following protocol. 
 We assume that the malicious nodes are controlled by an abstract entity that we call the \textit{attacker}. We assume that the attackers are computationally limited and cannot break the signature schemes or the cryptographic hash functions involved.  
 
In classic consensus protocols, the communication model already covers the adversary behaviors, as delaying blocks is essentially the only way an attacker can influence the system.
This is no longer true for the Tangle 2.0 protocol and we focus in this paper on attacks  on the voting layer.
In these kinds of attacks, the  attacker is in possession of some proportion of voting power, i.e. weight $q$, and uses this weight to manipulate the votes on the Tangle.
We are interested in the problem of metastability, which aims to affect liveness and, in extreme cases, safety.
The strategy of the attacker is then to keep the honest nodes in an undecided state.

\subsection{Bait-and-Switch Attack}

We implement several types of adversary strategies from  \cite{OTV} in the  simulator \cite{otv-simulator}. However,  we focus here on the Bait-and-Switch attack that seems to be the most effective strategy to attack liveness in the Tangle 2.0 protocol. Moreover, this attack is shown in theory to be effective even when the attacker has no control over the communication layer.

In contrast to typical balancing attacks, the Bait-and-Switch attack relies less on keeping the conflict weights symmetrical, but rather the attacker makes the honest nodes chase the ever-changing heaviest (measured in AW) transaction.  The attack seems to be most effective in situations where the adversary has the largest weight among all nodes.

The attack starts with the adversary issuing  a pair of conflicting transactions on which honest nodes start to vote. 
Then, before any of the two options accumulate more weight than the adversary has, the adversary issues another transaction conflicting with the previous ones, making it the heaviest transaction (measured in AW). The honest nodes will adapt their opinion to the newly issued conflict, but then the adversary issues another transaction conflicting with the previous ones and supports it with his weight. Thus, the new conflicting transaction becomes the heaviest in AW. This process can be repeated indefinitely unless a certain transaction accumulates more approval weight than the adversary has.

\section{Simulation Components}
\label{sec:simulation}

To achieve efficient exploration of the Tangle 2.0 protocol, only the necessary components are implemented in the simulator, and some of them are simplified. In this section, we explain the simplifications of the simulator and the reasons for them.

\subsection{Conflicts and Colours}
    \begin{figure}
        \centering
        \includegraphics[width=0.45\textwidth                  ]{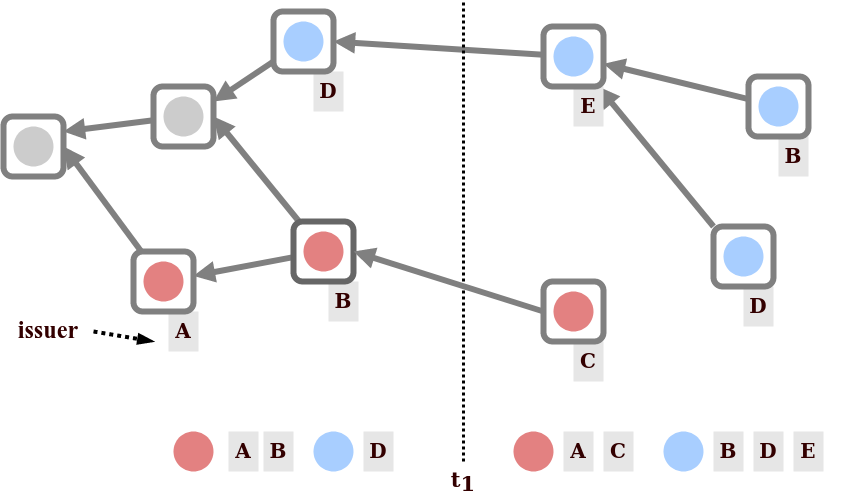}
        \caption{Tracking of the supporters for AW calculation.}
        \label{fig:colors}
    \end{figure}
We encode conflicting transactions with different colors. For instance, Figure \ref{fig:colors} presents the Tangle with two conflicting transactions: the red and blue colors are introduced in a block by nodes $A$ and $D$, respectively. 

Considering the Tangle state until time $t_1$ separated by the dashed line, we can count the supporters for each branch by summing up the weights of block issuers in the future cone of the block that introduced each branch. Thus for $t<t_1$, red is supported by nodes $A$ and $B$, and for  blue by node $D$. After some time, the support changes. Notice that node $B$ changed its opinion and attaches to the blue part of the Tangle, such that its previous vote for the red branch is canceled.

We want to note that  two types of references are introduced in \cite{OTV}, i.e., block reference and transaction reference. Here we consider only the basic type of reference, which is a block reference. Thus, it is not allowed to reference blocks that belong to conflicting branches (i.e., they have different colors).

\subsection {Communication layer}
\label{sec:color-voting}
The simulated environment reflects a situation in which network participants are connected in a peer-to-peer network, where each node has  $k$ neighbors. Nodes can gossip, receive blocks, request for missing blocks, and state their opinions whenever conflicts occur. In order  to fully control the network topology, we produce a peer-to-peer network using a Watts-Strogatz network \cite{Watts1998collective}. 
In order to mimic a real-world behavior, the simulator allows specifying the network delay and packet loss for each node's connection.

\subsection{Agent-based node model}

In the simulation, nodes act as different independent agents asynchronously. This means that different nodes can have different perceptions of the Tangle at any given moment of time. Thus, nodes have their own local Tangles for calculating the approval weights and performing the Tangle 2.0 protocol. 

We assume that the number of nodes does not change during the simulation period, and all the honest nodes are actively participating in the consensus mechanism. With this assumption, all nodes can be considered active nodes (i.e., their weight is accounted for during the AW calculations). 

\subsection{Weight Distribution}

As explained in Section \ref{sec:reputation} a scarce resource is assumed as a Sybil protection mechanism.  We also consider that the node's weight is derived from or related to the underlying source token, which is the case in IOTA. To model the weight of the nodes in the network we use the Zipf empirical law, which is proven to govern an asymptotic distribution of weights near its upper tail \cite{zipfs-law}. In \cite{muller2021} it was shown that in the example of $100$ IOTA's richest nodes, they follow approximately the Zipf law with $s=0.9$. 
Moreover, the Zipf law has the  advantage, i.e., by changing the parameter $s$, to model the network behavior for varying degrees of (de)centralization. For example, a homogeneous situation is modeled with  $s=0$, while for $s$ above $1$ a more centralized weight distribution can be investigated.
Note that, in the simulator, nodes' weights are constant over time, and thus there is no concept of weight delegation or changes in the weight distribution caused by transactions and events happening in the network.

\subsection{SRRS Implementation}
\label{sec:srrs-implementation}
If the conflict cannot be resolved until time $D\textsubscript{start}$, then at the time of $D\textsubscript{start}$ and at the end of each epoch, every node will calculate a hash based on the opinion of the node and the random number $x$. 

In our simulations, the random variable $x$ is  uniformly distributed on the interval [0.5, 0.66], where 0.66 is the value of the confirmation threshold $\theta$. The hash is calculated as the SHA-256 of $y$, where $y$ = $o$ + $c$ * $x$, and $c$ is a predefined constant value. $c$ is set to be $1000$ in our experiments. 
Every node compares the AW of the different conflicting transactions with the random threshold $x$. If one of the conflicting transactions has more AW than $x$ it will vote for this transaction. Otherwise, it will vote in the next $D$ period for the transaction with the minimal hash described above. 
This process keeps running periodically until all honest nodes decide on the same transactions or the maximal simulation time of $60$s is reached.

\section{Simulation Setup}
\label{sec:setup}
\begin{table}[]
    \centering
    \caption{Default Experimental Settings}
    \begin{tabular}{|c|l|c|}
         \textbf{Symbol} & \textbf{Description} & \textbf{Value} \\
         $N$ & Node count & 100 \\
         $s$ & Zipf parameter of weight distribution & 0.9 \\
         $\theta$ & Confirmation threshold & 66\% \\
         $n_p$ & Parents count & 8 \\
         BPS & Blocks per second (BPS) & 100 \\
         $l$ & Packet loss & 0 \\
         $\gamma$ & Watts-Strogatz rewiring probability & 100\% \\
         $k$ & Number of neighbors & 8 \\
         $q$ & Adversary weight  & 5\% \\
         $d$ & Network delay & 100ms \\
         $t\textsubscript{max}$ & Maximum simulation time & 60s \\
         
         \end{tabular}
    \label{tab:default_settings}
\end{table}

Table \ref{tab:default_settings} lists the default experiment setup in our simulations. The node count is the number of nodes in the network. Each node has a weight, which influences its impact on the algorithm. The average block issuance rate of a node is assumed to be proportional to the weight.
In addition, we assume the block issuance time interval of nodes follows a Poisson distribution \cite{Penzkofer2021ImpactOD}. The default confirmation threshold is set to be $\theta=66\%$.

The  parents' count (or the number of references) is set to be $n_p=8$. 
These parents are chosen randomly (with replacement) among all visible tips.
The throughput is measured as the total number of blocks per second (BPS) of all nodes combined; its  default value is BPS=$100 s^{-1}$. 
The packet loss is set to be $l=0$, i.e., no packets are lost. The minimum and maximum delays are the upper and lower latency bounds of packets between each pair of peers in the network, respectively. Both in the honest and adverse environments, the delay from an honest node or an adversary is set to be $d=100ms$, which is the packet delay transmitted from one node to its neighbors.
The default Watts-Strogatz rewiring probability \cite{Watts1998collective} is set to be $\gamma=1$. In addition, the neighbor count is the number of neighbors of a node in the Watts-Strogatz topology, whose default value is $8$. 
$t\textsubscript{max}$ denotes the maximum simulation time, and the default value is set to be $t\textsubscript{max}=60s$. 
The simulation will be terminated automatically and regarded as a security failure if the conflicts cannot be resolved before the predefined $t\textsubscript{max}$.

\begin{figure}
    \centering
    \includegraphics[width=0.5\textwidth]{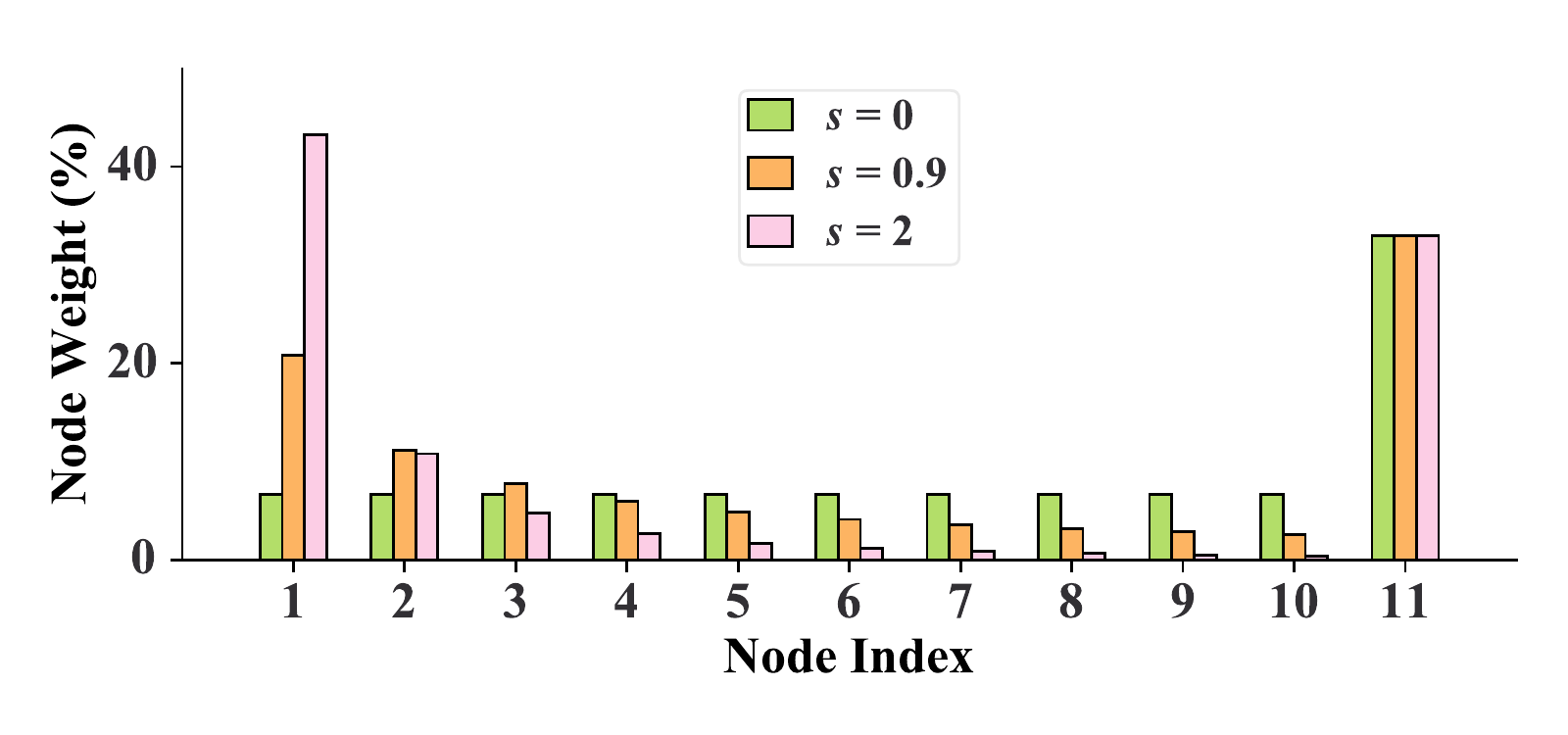}
    \caption{Node weight distribution example with $q=33\%$, $N=11$. The node index of the adversary is 11.}
    \label{fig:node_weight_example}
\end{figure}

The adversary weight is set to be $q$, which is the percentage of the total weight controlled by the adversary node. Figure \ref{fig:node_weight_example} shows weight distribution examples with $q=33\%$, $N=11$ for several values of $s$. In the figure, the index of the adversary node is $11$, and nodes with indices $1$ to $10$ are honest nodes. As $33\%$ of the total weight is occupied by the adversary, the remaining $67\%$ of the total weight is distributed among the $10$ honest nodes.

Based on the different settings of the above parameters, the robustness and efficiency of the Tangle 2.0 can be explored and simulated thoroughly.

\section{Simulation Results}
\label{sec:results}

To demonstrate the robustness of the Tangle 2.0, we first explore the consensus time distributions of the OTV protocol under adversary environments. The consensus time is defined as the time period measured from the time that a double spend happens to the time that all the honest nodes decided on the same opinion, i.e., reach  consensus. By measuring the consensus time, the safety and liveness of the protocol can be analyzed. Next, we analyze the confirmation time in expected scenarios (i.e., $N=100$ and $s=0.9$). The confirmation time is measured in the local Tangles of all honest nodes and is defined as the time period from the block issuance time to the time when its cumulative AW $>\theta$. By measuring the confirmation time distributions, the liveness of the protocol is analyzed. In the end, we show  experimental results that non-conflicting blocks are immune to attacks from adversaries.

\begin{figure}
     \centering
     \begin{subfigure}[b]{0.5\textwidth}
         \centering
         \includegraphics[width=1\textwidth]{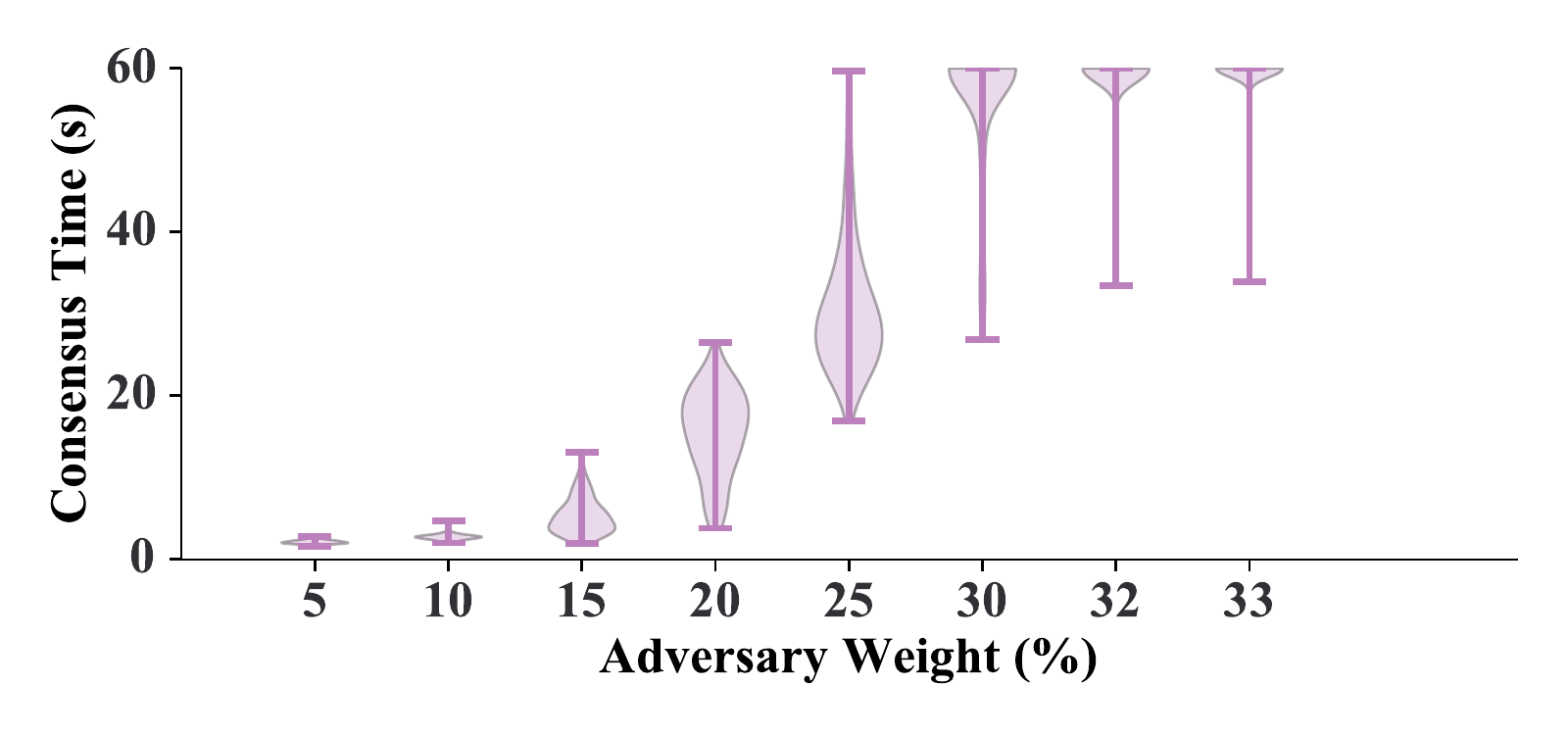}
         \caption{$s=0$}
         \label{fig:bs_wo_srrs_s0}
     \end{subfigure}
     \vfill
     \begin{subfigure}[b]{0.5\textwidth}
         \includegraphics[width=1\textwidth]{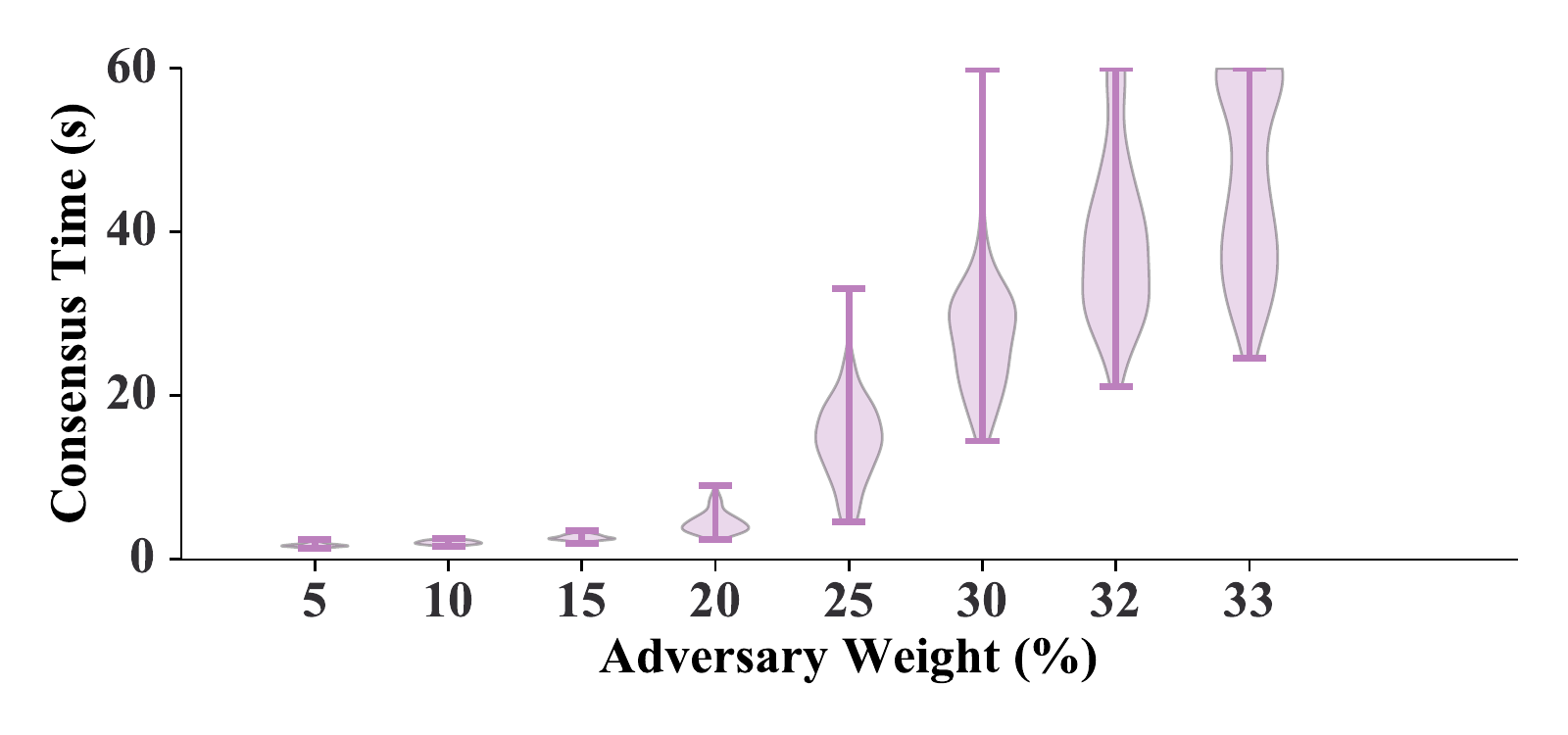}
         \caption{$s=0.9$}
         \label{fig:bs_wo_srrs_s0.9}
     \end{subfigure}
     \vfill
     \begin{subfigure}[b]{0.5\textwidth}
         \includegraphics[width=1\textwidth]{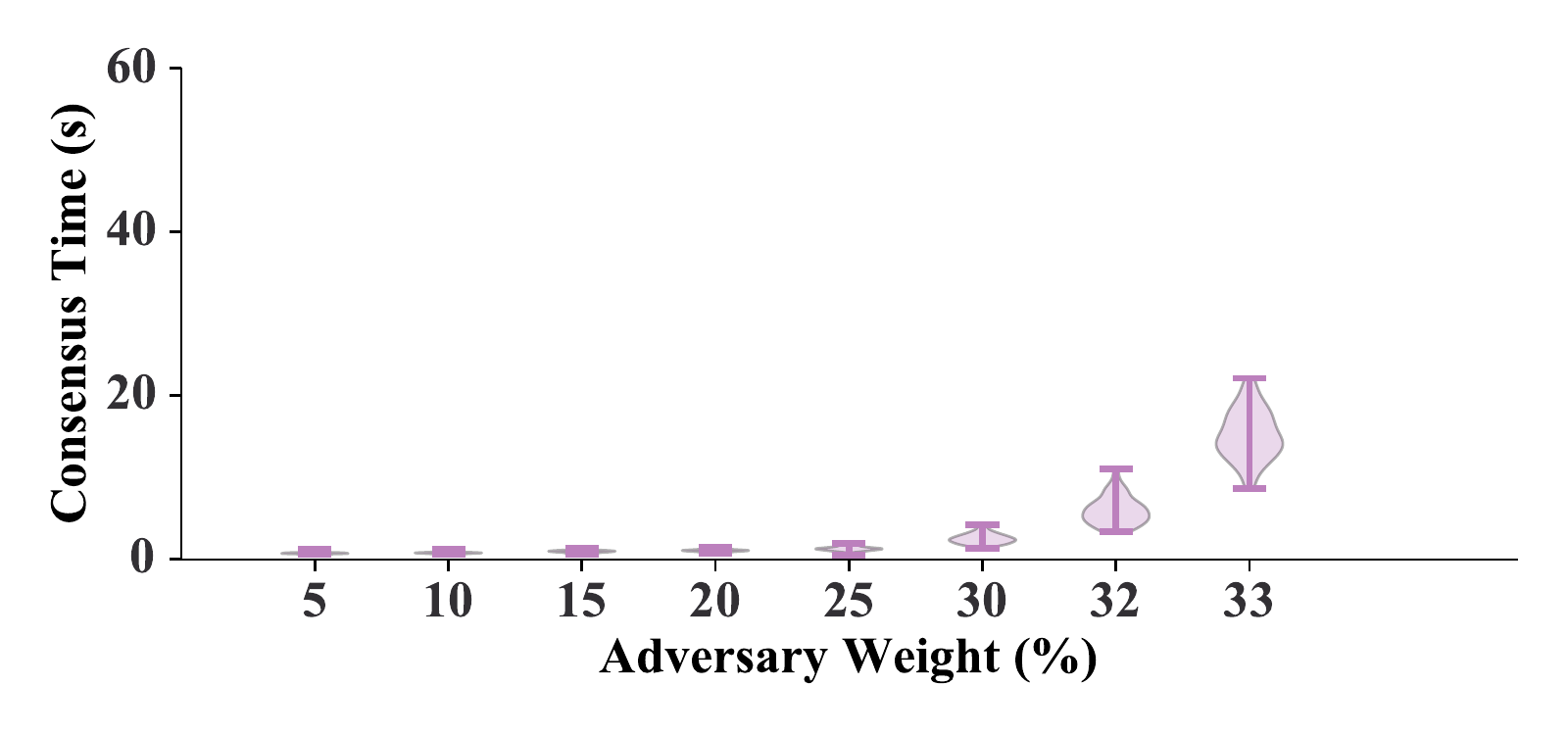}
         \caption{$s=2$}
         \label{fig:bs_wo_srrs_s2}
     \end{subfigure}
        \caption{OTV consensus time distributions under Bait-and-Switch attack, without SRRS ($N=100$).}
        \label{fig:bs_wo_srrs}
\end{figure}

\subsection{OTV Protocol without SRRS}\label{sec:withourSRSS}

Figure \ref{fig:bs_wo_srrs} shows the experimental results of the OTV consensus time distributions under the Bait-and-Switch attack and without SRRS. Different adversary weights, ranging from 5\% to 33\%, and Zipf distributions, $s\in\{0, 0.9, 2\}$, are simulated. Each experimental setting is simulated $100$ times. For each simulation run, the consensus time of the double spend  is collected. As Fig. \ref{fig:bs_wo_srrs_s0} shows, when $q \leq 20\%$, all of the $100$ simulation runs can successfully resist the Bait-and-Switch attack, where all the conflicts can be resolved within $30$s. When the adversary weight $q$ is increased to $\geq 25\%$, liveness failures occur, i.e., conflicts are failed to be resolved within $60$s.

With increased $s$, honest nodes are more likely to follow the opinions of the honest nodes with heavy weight, which makes the conflicts easier to be resolved.
We can observe this in Fig.  \ref{fig:bs_wo_srrs_s0.9}, where the Zipf parameter is increased to $s=0.9$. With the higher value of $s$ no liveness failures occur for $q \leq 25\%$. Also, most of the runs can resist the Bait-and-Switch attack for $q=30\%$. 

Figure \ref{fig:bs_wo_srrs_s2} shows the results of $s=2$ under the same attack. For all the cases shown, even $q$=33\%, no liveness failures occur. This provides strong evidence of the robustness of the OTV protocol, in a more centralized setting.

\subsection{OTV Protocol with SRRS }

In this section, we add the SRRS on top of the OTV protocol, and further investigate the robustness of the Tangle 2.0.

\begin{figure}
     \centering
     \begin{subfigure}[b]{0.5\textwidth}
         \centering
         \includegraphics[width=1\textwidth]{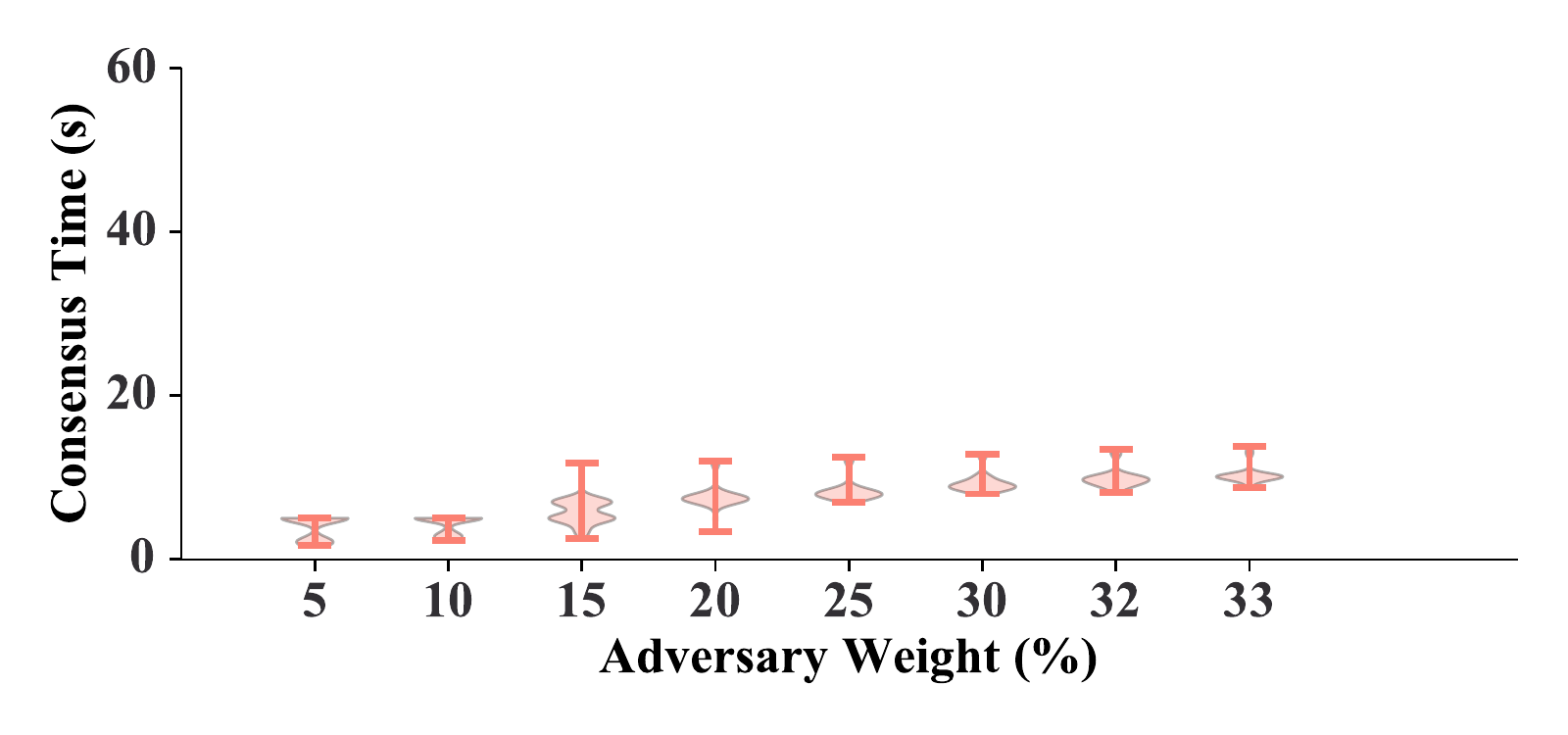}
         \caption{$s=0$}
         \label{fig:bs_w_srrs_s0}
     \end{subfigure}
     \vfill
     \begin{subfigure}[b]{0.5\textwidth}
         \includegraphics[width=1\textwidth]{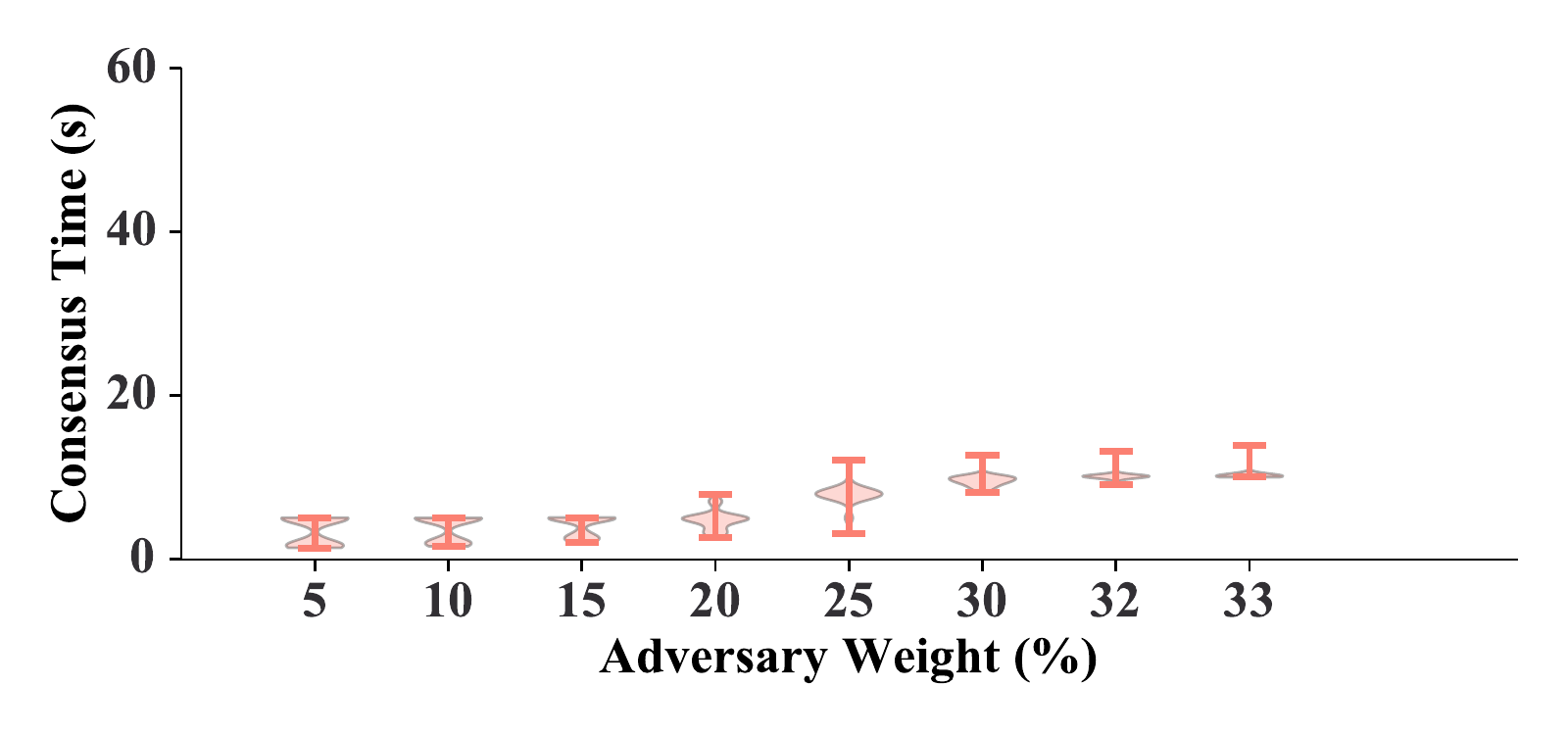}
         \caption{$s=0.9$}
         \label{fig:bs_w_srrs_s0.9}
     \end{subfigure}
     \vfill
     \begin{subfigure}[b]{0.5\textwidth}
         \includegraphics[width=1\textwidth]{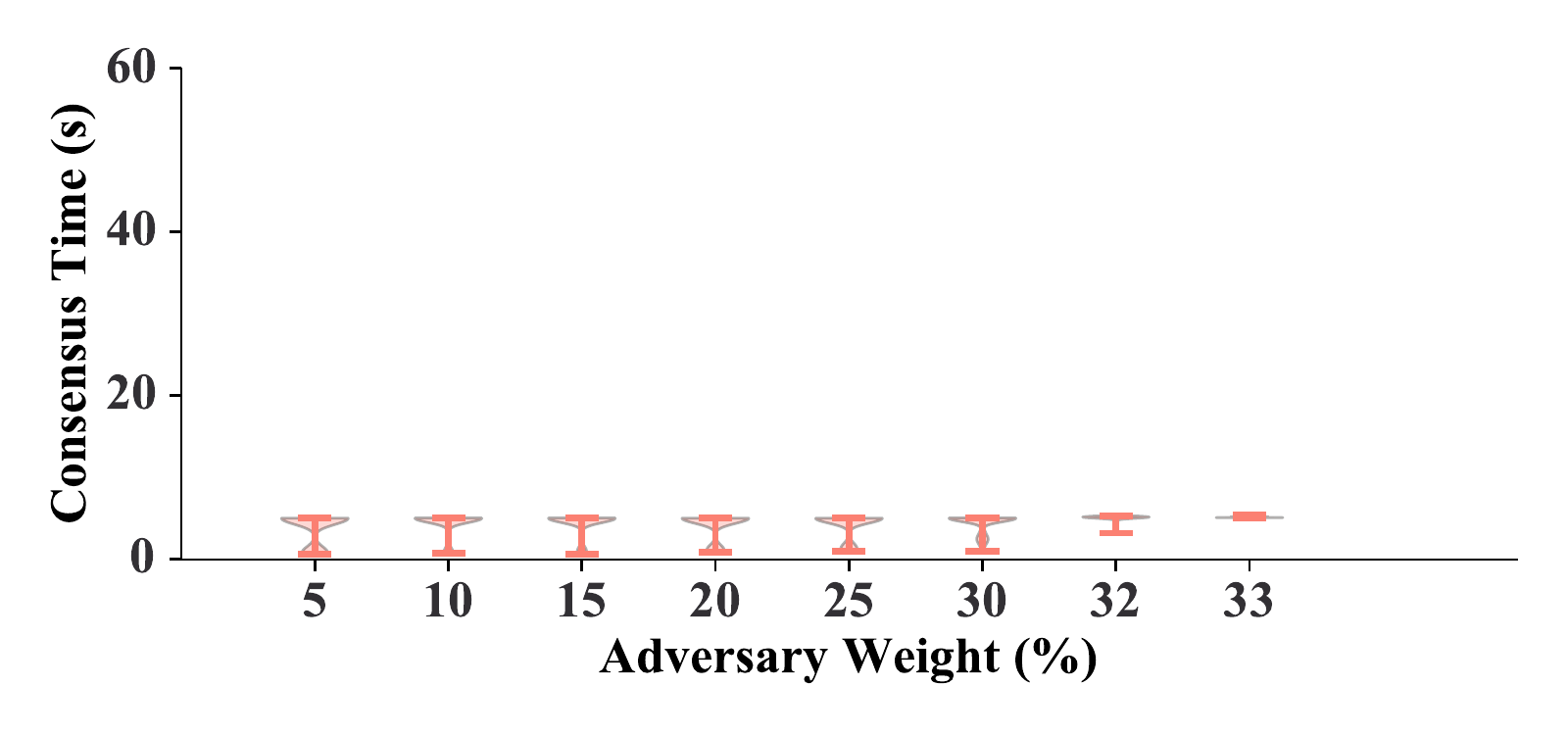}
         \caption{$s=2$}
         \label{fig:bs_w_srrs_s2}
     \end{subfigure}
        \caption{OTV consensus time distributions under Bait-and-Switch attack, with SRRS ($N=100$).}
        \label{fig:bs_w_srrs}
\end{figure}

Figure \ref{fig:bs_w_srrs} shows the experimental results of the consensus time distributions under the Bait-and-Switch attack, and with SRRS. The parameter variations of the adversary weights and Zipf distributions are the same as those in Section \ref{sec:withourSRSS}. As shown in the Fig. \ref{fig:bs_w_srrs}, no security failures occur in all the simulations. For all the cases, conflicts are able to be resolved within $20$s. Furthermore, for $s=2$, as shown in Fig. \ref{fig:bs_w_srrs_s2}, all the conflicts are able to be resolved within $7$s. Based on the simulation results, the Tangle 2.0 is able to resist the Bait-and-Switch attacks, even in extreme cases, where all the honest nodes have equal weight, and the adversary node owns the upper limit of the theoretical weight.

\subsection{Confirmation Time Analysis}

In this section, we analyze the confirmation times of honest transactions in an honest and adverse environment.

\subsubsection{Healthy Environment}

\begin{figure}
     \centering
     \includegraphics[width=0.5\textwidth]{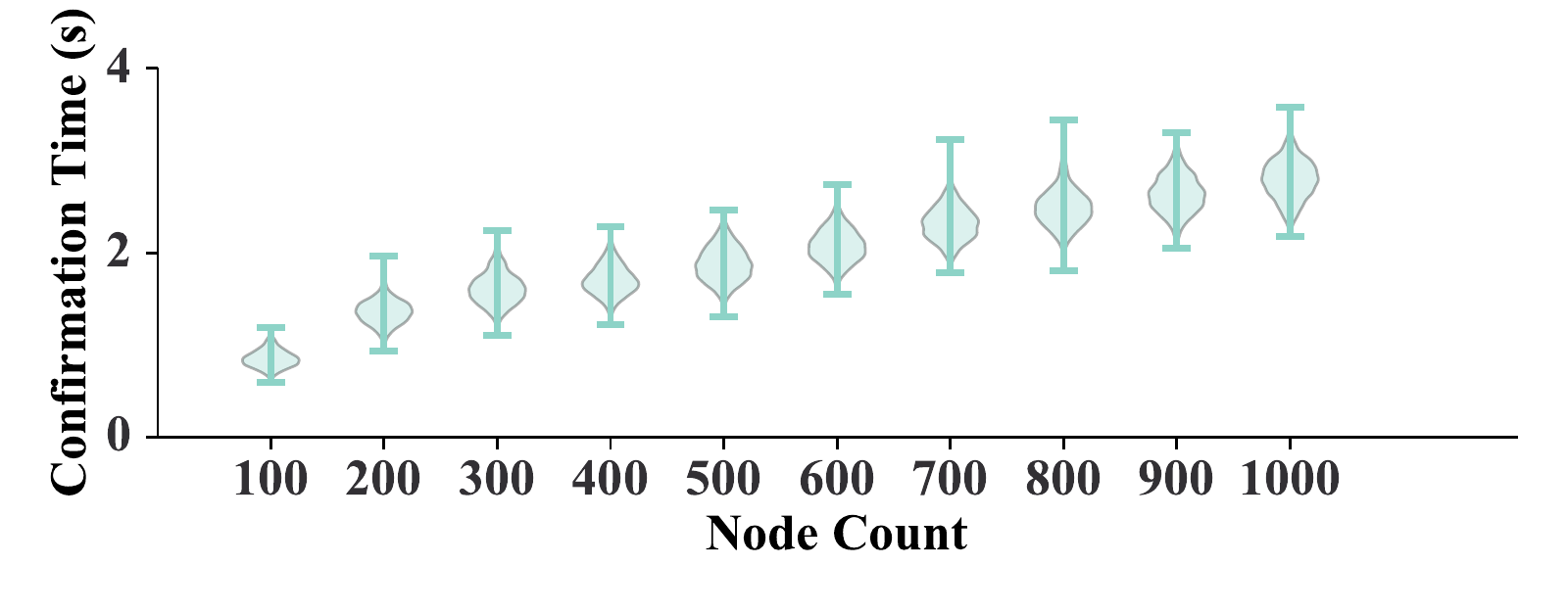}
     \caption{Confirmation time distributions with different $N$'s ($s=0.9$).}
     \label{fig:ct_different_n}
\end{figure}

To address the scalability of the protocol we investigate the confirmation time distribution in an honest  environment, i.e. $q=0$, for different node counts. As Fig. \ref{fig:ct_different_n} shows, the confirmation time increases monotonically with increasing node counts. The reason is that when the node count increases, the weights of nodes will be distributed across more nodes, and thus more time is required for a block to be approved by the necessary amount of weights.

\begin{figure}
     \includegraphics[width=0.5\textwidth]{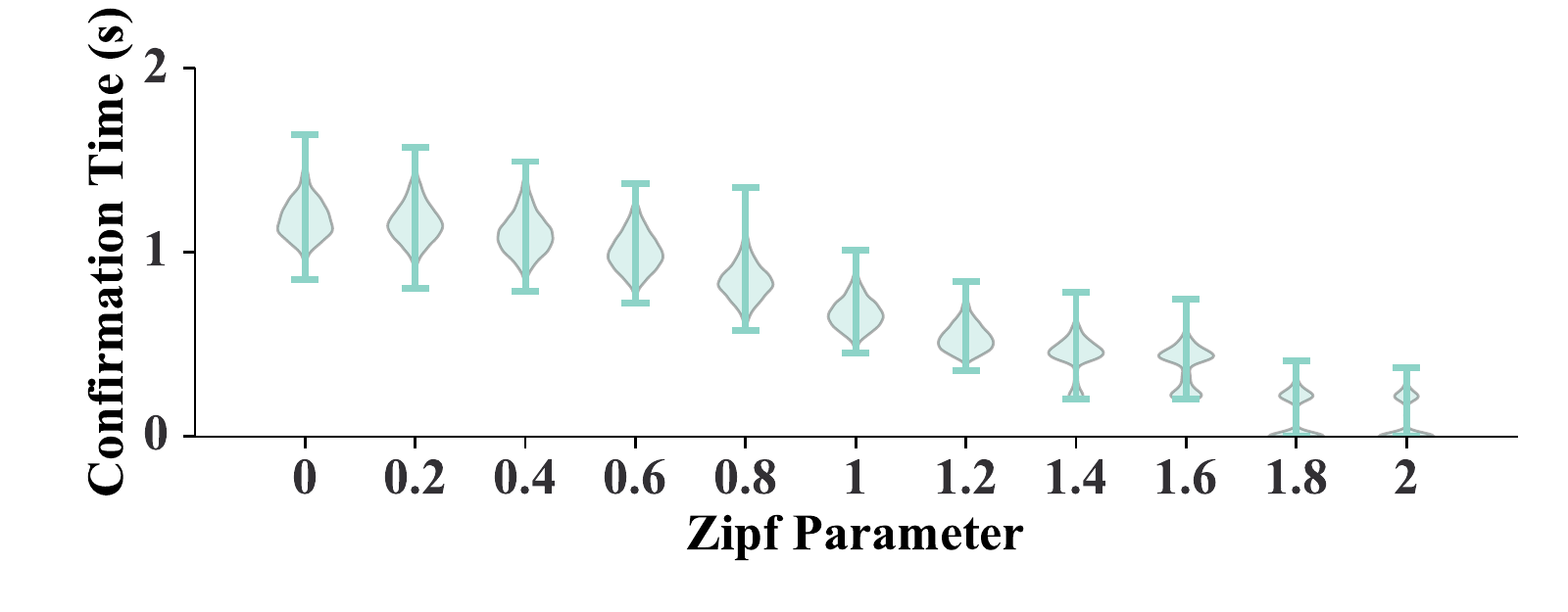}
     \caption{Confirmation time distributions for different Zipf parameters $s$ ($N=100$).}
     \label{fig:ct_different_s}
\end{figure}

To analyze the performance with the degree of decentralization of the network, Fig. \ref{fig:ct_different_s} shows the confirmation time distributions for several Zipf parameters $s$ with $N=100$. When the parameter $s$ decreases, the weights of nodes are distributed more evenly, and a given block needs to be approved by more nodes, which consequently leads to longer confirmation times. As Fig. \ref{fig:ct_different_s} shows, for an extreme case where the Zipf parameter equals $0$, all the blocks can still be confirmed within $2$s.

\begin{figure}
     \centering
     \includegraphics[width=0.5\textwidth]{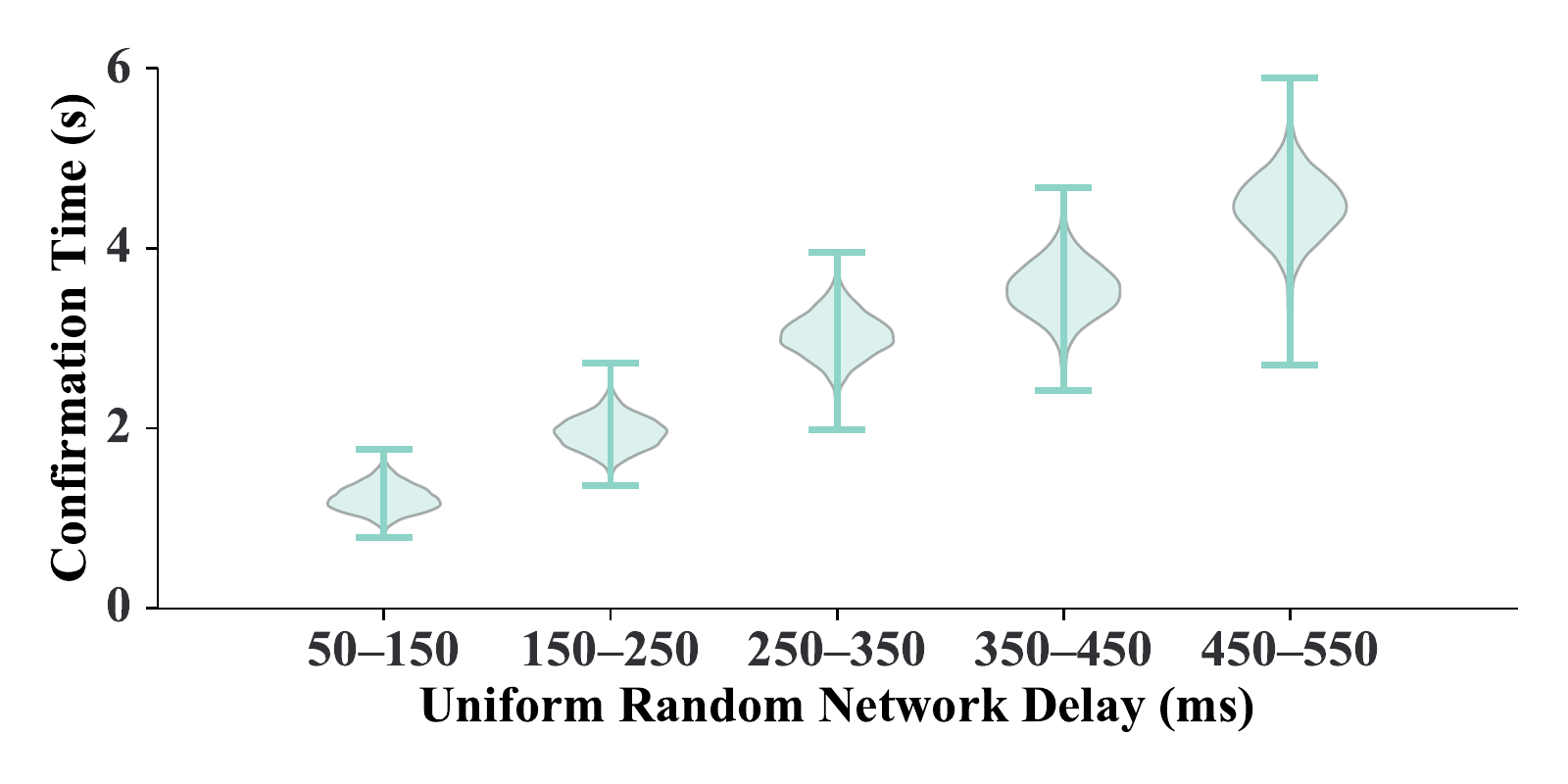}
     \caption{Confirmation time distributions with different uniform random network delays ($N=100$ and $s=0.9$).}
     \label{fig:ct_different_delay}
\end{figure}

Figure \ref{fig:ct_different_delay} shows the confirmation time distributions under different uniform random network delays. The confirmation time becomes longer monotonically with a longer network delay. Note that the delay for a modern network is around $100$ms, where $\leq{2}$s confirmation time can be achieved in the protocol.

\begin{figure}
     \centering
     \includegraphics[width=0.5\textwidth]{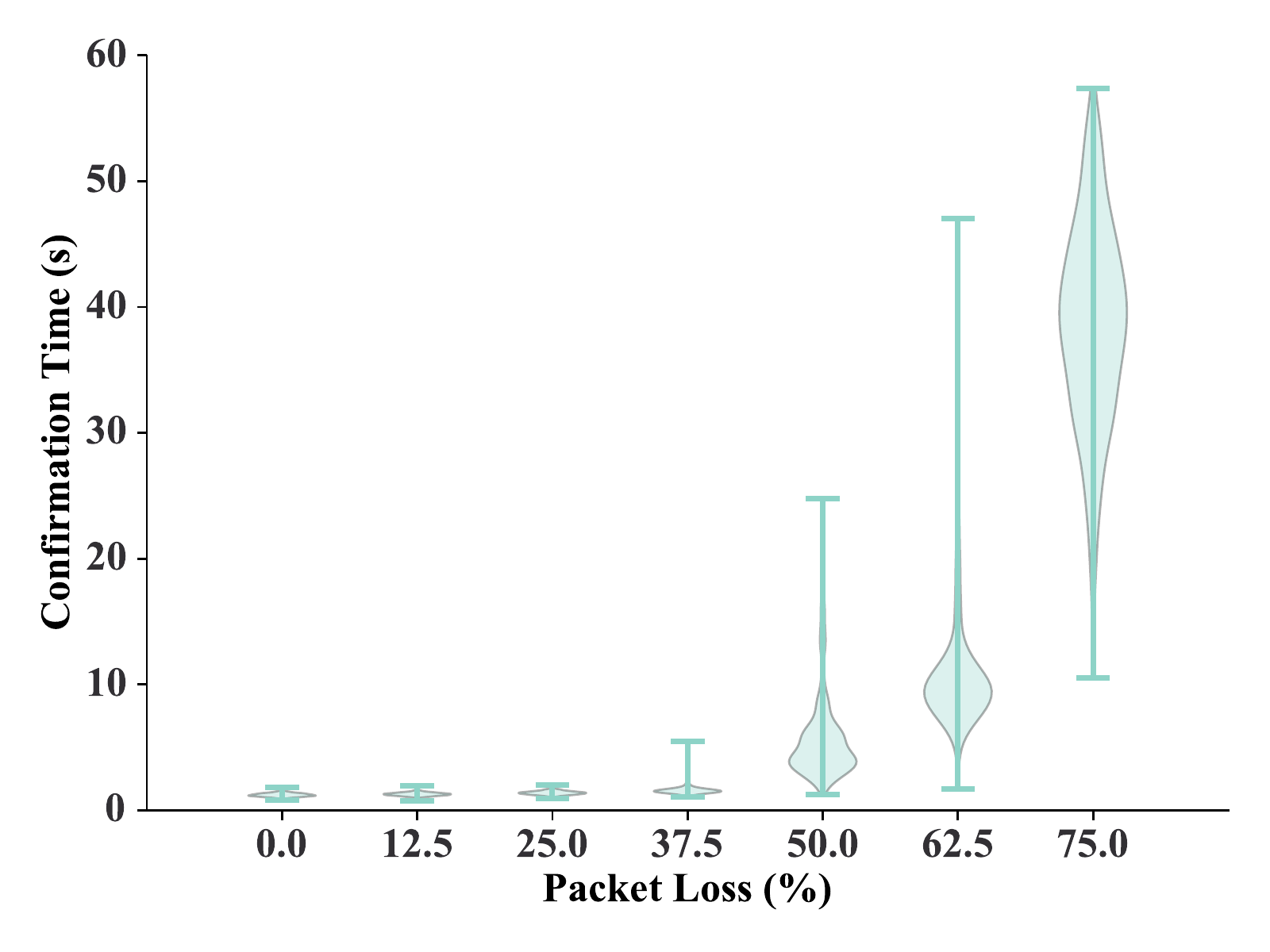}
     \caption{Confirmation time distributions with different packet losses ($N=100$ and $s=0.9$).}
     \label{fig:ct_different_package_loss}
\end{figure}

In a real network environment, a packet might get lost due to network congestion and/or software/hardware issues. This will introduce the packet failing to transmit from one peer to another. In the network layer of the IOTA protocol, a \textit{solidification process} is implemented. In the process, each node maintains a local tangle for computing the approval weights of transactions by tangle traversing. When a missing transaction is identified in the traversing (i.e., the parent(s) of a transaction is missing), the node will request the missing transaction from other peers every $5$s, until the transaction is received and contained in its local tangle. Figure \ref{fig:ct_different_package_loss} shows confirmation time distributions with different packet losses. Thanks to the solidification process, the confirmation time can still remain $\leq{2}$s when the packet loss is $\leq{25\%}$ for $s=0.9$ and $N=100$.

\subsubsection{Adverse Environment}

\begin{figure}
     \centering
     \includegraphics[width=0.5\textwidth]{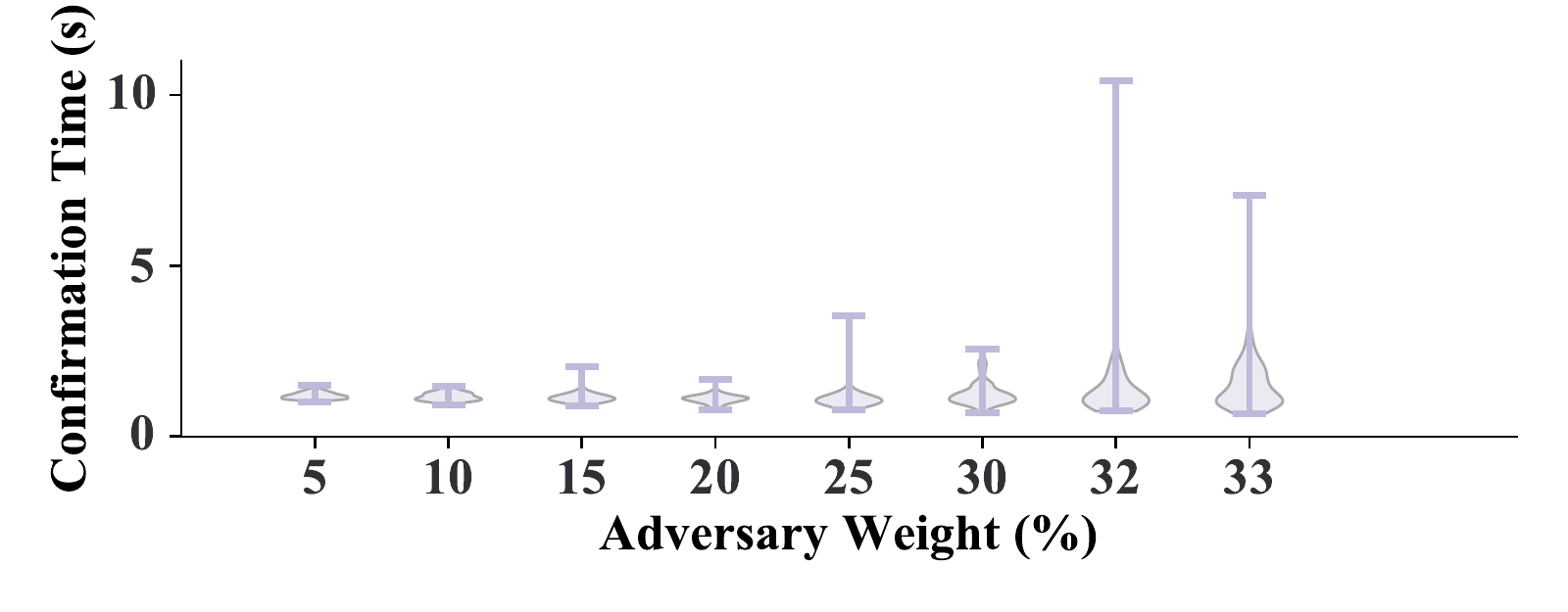}
     \caption{Confirmation time distributions of non-conflicting transactions with $s=0.9$ and $N=100$.}
     \label{fig:ct_bait_and_swtich}
\end{figure}

Figure \ref{fig:ct_bait_and_swtich} shows the confirmation time distributions of non-conflicting transactions under the Bait-and-Switch attack, where the Zipf parameter and the node count are $s=0.9$ and $N=100$, respectively. As shown in the figure, most of the non-conflicting blocks can be confirmed within $1$ to $2$s, which are not affected by the presence of conflicts. Note that there are a few outliers. These can be explained by non-conflicting transactions (i.e., the outliers) being  orphaned  because the voting powers of the honest nodes are split to vote for the different conflicting transactions before the conflict resolution.
In the case of  an orphanage,  the issuers of the affected block reattaches the transaction in a new block after $5$s. This is possible, because the issuance of different blocks with the same transaction is not considered a double spend but rather, on the UTXO ledger level, is considered as the same transaction. 
In the IOTA 2.0 protocol, this liveness problem for honest  transactions is resolved more elegantly  by the introduction of the second type of reference, i.e., a transaction reference \cite{OTV}. In addition, due to the fixed confirmation threshold, which is $\theta=66\%$, as the adversary occupies more weight, non-conflicting transactions might wait for a longer time to accumulate their AW from different honest nodes.

\section{Conclusion}
\label{sec:conclusion}
In this paper, we gave a first performance analysis of the Tangle 2.0 protocol. The confirmation time distributions based on different environments and under attacks are analyzed. The experimental results show that the Tangle 2.0 can resist the Bait-and-Switch attack and achieve a short consensus time, even in extremely adverse environments. 
Moreover, in common scenarios, the protocol can achieve  confirmation time in the order of seconds, and confirmation times of most non-conflicting transactions are not affected noticeably by the presence of conflicts.

\ifCLASSOPTIONcompsoc
  \section*{Acknowledgments}
\else
  \section*{Acknowledgment}
\fi

The authors would like to thank the developer team of the GoShimmer software, to support this study with the prototype implementation of the Tangle 2.0 protocol.

\bibliographystyle{IEEEtran}
\bibliography{bibliography}
\end{document}